\documentstyle[aps,epsf]{revtex}
\begin{document}
\begin{flushright}
IHEP 96-41\\ 
%May 1996\\
%hep-ph/9605451
\end{flushright}
\begin{center}
{\large\bf Hard approximation in two-particle hadronic decays of $B_c$
at large recoils}\\
\vspace*{3mm}
V.V.~Kiselev\\
Institute for High Energy Physics,\\
Protvino, Moscow Region, 142284, Russia\\
E-mail: kiselev@mx.ihep.su
\end{center}
\begin{abstract}
{The two-particle decays of $B_c^+\to \psi \pi^+(\rho^+)$ and
$B_c^+\to \eta_c \pi^+(\rho^+)$ are considered in a way taking into 
account a soft binding of quarks in the heavy quarkonia and a hard 
gluon exchange between the constituents at large recoil momenta of 
$\psi(\eta_c)$. An approximate double enhancement of the amplitudes is found
because of the nonspectator $t$-channel contribution.}
\end{abstract}
\vspace*{3mm}

\section{Introduction}

The QCD dynamics plays a significant role in an extraction of the electroweak
theory parameters in the heavy quark sector. One of the systems allowing one 
to perform an exact numerical study of the heavy quark interactions, is
the $(\bar b c)$ system, the heavy quarkonium with the mixed flavor.
At present, the experimental search for the $B_c$ meson, the basic
pseudoscalar $1S$-state of the $(\bar b c)$ system, is curried out at
CDF \cite{1} and ALEPH \cite{2}. 

General properties of the $B_c$ meson
family can be quite reliably predicted in the theoretical investigations
allowing one to make an objective experimental search for the $B_c$ 
observation (see the review on the $B_c$ physics in \cite{3}).
The spectroscopic characteristics of $(\bar b c)$ family can be calculated
in the framework of phenomenological nonrelativistic potential models
\cite{4,5} and their relativistic modifications \cite{6}. The strong and
electromagnetic interactions conserving the flavor, do not give the 
annihilation modes of the $(\bar b c)$ state decays. Therefore, the
excited levels radiatively transform into the lowest longliving pseudoscalar 
$B_c^+$ state decaying due to the weak interaction. The mass of this state,
$M(B_c^+) = 6.25\pm 0.03$ GeV, and its leptonic constant, $f_{B_c} =
385\pm 25$ MeV, can be predicted in the framework of potential models
\cite{4,5}, QCD sum rules \cite{7,8,13} and in the lattice computations 
\cite{9}.
The life time, $\tau(B_c) = 0.55\pm 0.15$ ps was estimated in several papers,
where one took into account corrections caused by the quark binding inside
the heavy quarkonium in two ways, the phenomenological one \cite{10} as well as
in the operator product expansion for the weak currents of decays of the
heavy quarks composing the $B_c$ meson \cite{11}.

From the viewpoint of the experimental selection of the $B_c$ meson signal 
in a hadronic background, the preferable modes for the $B_c$ observation are
those, wherein the final state contains the $\psi$ particle, which
can be reliably identified in the leptonic decay, $\psi\to l^+l^-$.
As $\bar c$-quark produced in the $\bar b\to \bar c W^{*+}$ transition,
can bind the spectator $c$-quark of the $B_c$ meson with a high probability
into the $(\bar c c)$ meson, the relative yield of $\psi$ particles in the
$B_c$ decays should be enhanced in comparison with the branching ratio of
the $B_{u,d}$ meson decay modes with $\psi$ in the final state. Indeed,
under the obtained theoretical estimates in the framework of phenomenological
models of the meson, one should expect ${\rm BR}(B_c^+\to \psi X)\sim 17$ \%,
which is much greater than ${\rm BR}(B_{u,d}\to \psi X)\sim 1$ \%.

As for the semileptonic decay mode of $B_c^+\to \psi l^+\nu$, estimates of
its width calculated within the potential models \cite{10,12} and in the
QCD sum rules \cite{13}, point out the essential discrepancy between 
results obtained in these two approaches (the QCD sum rule estimate of the
$B_c^+\to \psi l^+\nu$ decay width is one order of magnitude less than
values given by the different models of heavy quarkonia). As was shown
in \cite{14}, this deviation can be removed, if one takes into account
the Coulomb corrections to the vertices of the meson quantum-number currents in
the framework of QCD sum rules. 

The semileptonic mode of $B_c$ decay
is suitable for the reliable experimental identification of the $B_c$ meson
at a rather large statistics of events with $B_c$ \cite{15}. However, at the
current experiments in $e^+e^-$-annihilation and hadron-hadron collisions, one 
has to expect the $B_c$ production rate, which evidently is not sufficient
to identify $B_c$ in the semileptonic mode \cite{16,17}. Therefore,
the two-particle decay of $B_c^+\to \psi \pi^+$ allowing one to find $B_c$
practically over a single event, is of the greatest interest in the 
experimental search for $B_c$. The estimate of its width calculated in the 
potential models, gives the branching fraction
$$
{\rm BR}^{\rm PM}(B_c^+\to \psi \pi^+)\approx 0.2 \%\;.
$$

However, in the semileptonic $B_c$ decay the region of low momenta for the
$\psi$ particle recoil dominates, and this allows one to apply the approximate
spin symmetry for the heavy $(\bar b c)$ and $(\bar c c)$ quarkonia \cite{18} 
and reliably to use the way of the transition form-factor calculation under 
the overlapping of the quarkonium wave functions. In contrast to the above 
transition, the two-particle modes of hadronic decays of
$B_c^+\to \psi \pi^+(\rho^+)$ and $B_c^+\to \eta_c \pi^+(\rho^+)$ require
a special consideration. This is related with the fact that at large momenta
of the recoil quark in the $\bar b \to \bar c \pi^+$ transition, the
$\bar c$-antiquark has to exchange by a hard gluon with the charmed $c$-quark
being in the initial state, to form  the bound $\psi(nS)$ or $\eta_c(nS)$
state in the region of low invariant masses of the $(\bar c c)$ pair due to
nonperturbative soft interactions of QCD. Thus, the feature of the two-particle
hadronic $B_c$ decays is determined by the fact that in the $\bar b$-quark
decays, the spectator quark is also heavy and, hence, at large energy release,
the description of exclusive production of the $(\bar c c)$ quarkonium in the
final state can not be performed in the framework of the spectator approach,
where the quark-spectator determines only the amplitude of a soft forming
of the bound state, so that the process of the hard weak decay can be 
factorized and it does not depend on the spectator. In the decays under
consideration, this spectator picture is not valid. So, one can use the 
Brodsky-Lepage hard scattering formalism \cite{bl}.

In this paper, we consider the exclusive $B_c^+\to \psi \pi^+(\rho^+)$ and
$B_c^+\to \eta_c \pi^+(\rho^+)$ decays in the framework of the hard
approximation at large recoils with taking into account the gluon exchange
to the $c$-quark in the initial state. In contrast to the spectator approach,
the hard $t$-channel exchange results in the approximate double enhancement of
the decay amplitudes, as it was recently  found  for $B_c^+\to \psi \pi^+$
\cite{19}.

In Section II we derive expressions for the amplitudes and widths of the
$B_c^+\to \psi \pi^+(\rho^+)$ and $B_c^+\to \eta_c \pi^+(\rho^+)$ decays and
compare them with the spectator formulae for the $\bar b\to \bar c \pi^+
(\rho^+)$ transitions. Numerical estimates of the decay widths are given 
in Section III, where theoretical uncertainties of the values are discussed.
The obtained results are summarized in the Conclusion.

\section{Calculation of two-particle widths of $B_c$}

In the framework of the nonrelativistic formalism for the heavy quark binding
into the $S$-wave quarkonium, we assume that the momentum of the quark,
composing the
meson, is equal to $p_Q^\mu = m_Q v^\mu$, where $v_\mu$ is the four-velocity of
quarkonium, so that the quarks inside the meson move with the same
four-velocity $v$. Moreover, the quark-meson vertex with nontruncated
quark lines corresponds to the spinor matrix
$$
\Gamma_V = \hat \epsilon\; \frac{1+\hat v}{2}\; 
\frac{\tilde f M_{nS}}{2\sqrt{3}}\;,
$$
for the vector quarkonium with $\epsilon_\mu$, being the polarization vector,
and
$$
\Gamma_P = \gamma_5\; \frac{1+\hat v}{2}\; 
\frac{\tilde f M_{nS}}{2\sqrt{3}}\;,
$$
for the pseudoscalar quarkonium. Here $M_{nS}$ is the $nS$-level mass and 
$\tilde f$ is related with the value of configuration wave function at the
origin
$$
\tilde f = \sqrt{\frac{12}{M_{nS}}}\; |\Psi_{nS}(0)|\;.
$$
The $\tilde f$ quantity can be related with the leptonic constants of states
\begin{eqnarray}
\langle 0| J_\mu(0)|V\rangle & = & i f_V M_V\; \epsilon_\mu\;, \nonumber\\
\langle 0| J_{5\mu}(0)|P\rangle & = & i f_P p_\mu\;, \nonumber
\end{eqnarray}
where $J_{\mu}(x)$ and $J_{5\mu}(x)$ are the vector and axial-vector currents
of the constituent quarks. Then the allowance for the hard gluon corrections
in the first order over $\alpha_s$ \cite{7,sv,bra,vol} results in
\begin{eqnarray}
\tilde f & = & f_V\; \bigg[1 - \frac{\alpha_s^H}{\pi}
\biggl(\frac{m_2-m_1}{m_2+m_1}\ln\frac{m_2}{m_1} -\frac{8}{3}\biggr)\bigg]\;,\\
\tilde f & = & f_P\; \bigg[1 - \frac{\alpha_s^H}{\pi}
\biggl(\frac{m_2-m_1}{m_2+m_1}\ln\frac{m_2}{m_1} - 2\biggr)\bigg]\;,
\end{eqnarray}
where $m_{1,2}$ are the masses of quarks composing the quarkonium. For the
vector currents of quarks with equal masses, the BLM procedure of the scale 
fixing in the "running" coupling constant of QCD \cite{blm} gives \cite{vol}
$$
\alpha_s^H = \alpha_s^{\overline{\rm MS}}(e^{-11/12}m_Q^2)\;.
$$
For the quarkonium with $m_1\neq m_2$, we assume
$$
\alpha_s^H = \alpha_s^{\overline{\rm MS}}(e^{-11/12}m_1 m_2)\;.
$$
Note, in the given estimates one considers the hard gluon corrections to
the quark-antiquark annihilation currents. The corresponding factors are
known exactly, and that is surprisingly, they can be obtained by 
the symbolic substitutions $m_1\to -m_1$, $V\leftrightarrow P$
from the exact expressions
for the hard gluon factors of the quark-to-quark transition currents
\cite{sv}, considered in HQET \cite{12+}, at 
the  prescription  of the absolute value for the logarithm argument. 
However, these substitutions do not 
lead to valid evaluations of the BLM scales determining $\alpha_s^H$.
The corresponding BLM scales in HQET have been calculated by M.~Neubert
\cite{12++}, and they do not give the exactly known result for the
quark-antiquark annihilation vector current \cite{vol}.

Further, the factor of the colour wave function $\delta^{ij}/\sqrt{3}$ stands
in the quark-meson vertex. 

The $\pi$ meson current corresponds to the 
axial-vector current of weak transition $A^\mu =f_\pi p_\pi^\mu$.
So, the given factorization  neglects  possible final state
interactions, which really seem to be small (see discussion in \cite{12-}).

\setlength{\unitlength}{1mm}
\begin{center}
\begin{picture}(150,60)
\put(-10,0){\epsfxsize=15cm \epsfbox{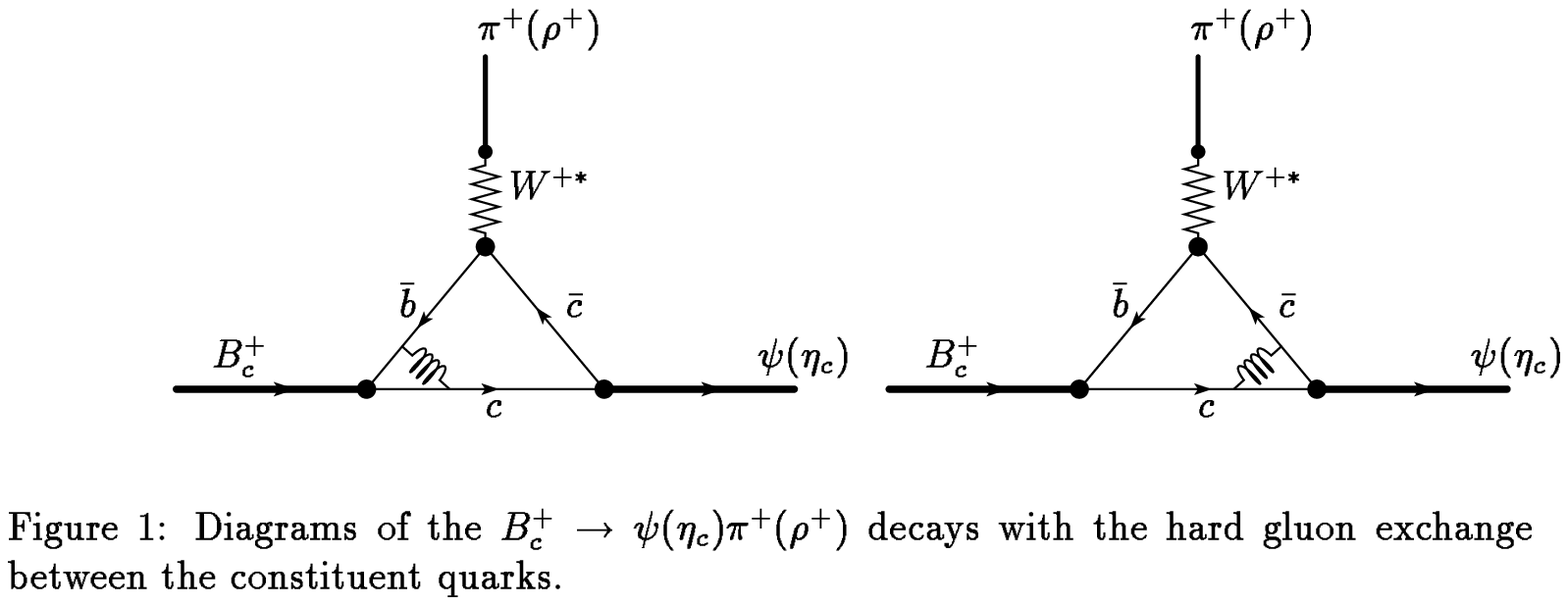}}
\end{picture}
\end{center}

Then the matrix elements of $B_c^+\to \psi \pi^+$ and $B_c^+\to \eta_c \pi^+$
decays calculated according to the diagrams in Fig.~1, take the form
\begin{eqnarray}
T(B_c^+\to \psi \pi^+) &=& \frac{G_F}{\sqrt{2}}\; V_{bc}\; \frac{32\pi\alpha_s}
{9}\; f_\pi \tilde f_{B_c}\tilde f_{\psi}\; \frac{M^2}
{m_\psi^2 (y-1)^2}\; (\epsilon \cdot v)\; a_1\;,
\label{t-}\\
T(B_c^+\to \eta_c \pi^+) &=& \frac{G_F}{\sqrt{2}}\; V_{bc}\; 
\frac{8\pi\alpha_s}{9}\; f_\pi \tilde f_{B_c}\tilde f_{\eta_c}\; 
\frac{(M^2-m^2_{\eta_c})\; (3M^2-2 M m_{\eta_c}+m^2_{\eta_c})}
{M m_{\eta_c}^3 (y-1)^2}\; a_1\;,
\label{t}
\end{eqnarray}
where $v$ is the four-velocity of $B_c$ meson, $M$ is its mass, $\epsilon$
is the polarization vector of $\psi$ particle, $m_\psi$ is its mass,
$y=v\cdot v_\psi$ is the product of the $B_c$ and $\psi$ four-velocities,
$$
y=\frac{M^2+m^2_\psi}{2M m_\psi}\;.
$$
In (\ref{t}) the notations for the $\eta_c$ state are analogous to the 
described ones. The $a_1$ factor is caused by the 
hard gluon corrections to the effective four-fermion weak interactions of
quarks. $a_1$ is evaluated for free quarks, so that $a_1= 1.22\pm 0.04$
\cite{3}. In present calculations, this correction is at the same level
as the hard corrections to the decay constants as well as the gluon propagator.
The value of the QCD coupling constant is determined by the gluon virtuality,
$k_g^2= - m^2_{\psi, \eta_c} (y-1)/2$, and it will be discussed in 
next Section.

The corresponding virtualities of $\bar c$
and $\bar b$ quarks, interacting with the hard gluon, are equal to
\begin{eqnarray}
k^2_c - m^2_c & = & 2k^2_g\;, \nonumber\\
k^2_b - m^2_b & = & 2 k^2_g M/m_{\psi, \eta_c}\;. \nonumber
\end{eqnarray}
Note, that  numerically the virtual $\bar c$-quark in the second
diagram in Fig.~1 is in the $t$-channel, since its four-momentum squared has a 
negative value. Therefore, one can see, that the corresponding contribution 
into the $B_c^+ \to \psi(\eta_c) \pi^+$ decay is 
definitely nonspectator, and the considered process is certainly hard.

From Eqs.(\ref{t-},\ref{t}) one gets the expressions for the total widths of
the $B_c^+\to \psi \pi^+$ and $B_c^+\to \eta_c \pi^+$ decays
\begin{eqnarray}
\Gamma(B_c^+\to \psi \pi^+) &=& G_F^2 |V_{bc}|^2\; \frac{128\pi\alpha_s^2}{81}
f^2_\pi \tilde f^2_{B_c}\tilde f^2_{\psi} \biggl(\frac{M+m_\psi}{M-m_\psi}
\biggr)^3\; \frac{M^3}{(M-m_\psi)^2 m_\psi^2}\; a_1^2,
\label{g-}\\
\Gamma(B_c^+\to \eta_c \pi^+) &=& \Gamma(B_c^+\to \psi \pi^+) \cdot
\frac{(3M^2-2 M m_{\eta_c}+m^2_{\eta_c})^2}{4 M^4}\;.
\label{g}
\end{eqnarray}

As for the analogous two-particle $B_c^+$ decays with $\rho^+$ in the final
state, one uses the approximate factorization of the transition current of the
virtual $W^{*+}$ boson into $\pi^+$ or $\rho^+$, and one finds that
the only difference between the squares of amplitudes for the pseudoscalar and 
vector states of the light quark systems, is the substitution of the 
quantity $f^2_\rho m^2_\rho (-g_{\mu\nu}+p^\rho_\mu p^\rho_\nu/m^2_\rho)$ 
instead of the $\pi^+$ meson current tensor $f^2_\pi p^\pi_\mu
p^\pi_\nu$. Then one can easily observe that after the summation over the
$\rho$ meson polarizations, the squares of the matrix elements coincide up to
the factor, so
\begin{equation}
\frac{\Gamma(B_c^+\to \psi\rho^+)}{\Gamma(B_c^+\to \psi\pi^+)} \approx
\frac{\Gamma(B_c^+\to \eta_c\rho^+)}{\Gamma(B_c^+\to \eta_c\pi^+)}
\approx \frac{f^2_\rho}{f^2_\pi}\;,
\label{rat}
\end{equation}
in the leading order over the small parameters, $m^2_\rho/m^2_c$,
$m^2_\rho/m^2_b$.

Write down the expressions for the spectator decays of $\bar b$-quark
\begin{eqnarray}
\Gamma(\bar b \to \bar c \pi^+) &=& G_F^2 |V_{bc}|^2\; \frac{m_b^3 f_\pi^2}
{16\pi}\; \biggl(1- \frac{m_c^2}{m_b^2}\biggr)^3\; a_1^2\;,
\label{spec-}\\
\Gamma(\bar b \to \bar c \rho^+) &=& \Gamma(\bar b \to \bar c \pi^+) \;
\frac{f^2_\rho}{f^2_\pi}\; \bigg\{1+\frac{m^2_\rho (m^2_b+m^2_c)}
{(m^2_b-m^2_c)^2}\bigg\}\;.
\label{spec}
\end{eqnarray}
From Eqs.(\ref{spec-},\ref{spec}) one can see that for the spectator decays
the relation between the yields of $\rho^+$ and $\pi^+$ mesons
$$
\frac{\Gamma(\bar b \to \bar c \rho^+)}{\Gamma(\bar b \to \bar c \pi^+)}
\approx \frac{f^2_\rho}{f^2_\pi}
$$
is valid in the leading approximation over the square of ratios of the $\rho$ 
meson mass over the heavy quark masses, as it takes place for the transitions
between the mesons (see (\ref{rat})). In the spectator decays the accuracy of 
the leading approximation used, is about 4 \%, that also points out the 
magnitude of the correction terms to relation (\ref{rat}) for the mesons.

The breaking of the spectator picture at large recoils in nonhadronic decays
of $B_c^+$ was also recently considered in \cite{cc}, where one studied
the $B_c^+\to D_s^+\gamma$ mode due to the flavor-changing neutral
current of $\bar b\to \bar s \gamma$. The discussion of the heavy quark 
symmetry at large recoils in the heavy-light meson transitions was given
in \cite{korner}, where the peaking approximation was introduced. This 
approximation corresponds to the quark-meson vertices used in this paper.

\section{Numerical estimates}

The accuracy of the given calculations is basically restricted by the
uncertainty in the choice of the QCD coupling constant value.
In Eqs.(\ref{t-},\ref{t}) $\alpha_s$ 
can be evaluated at the scale typical for the 
charm quark physics $\alpha_s\approx 0.30$. The higher order corrections
are beyond the scope of this paper. Nevertheless, to evaluate the
possible value of these corrections, one can use the BLM procedure
including the light quark loops in the virtual gluon propagator. So,
$\alpha_s$ is given at the scale of the gluon virtuality
\begin{equation}
k^2_g = - m^2_{\psi, \eta_c} (y-1)/2 \approx -1.2\; \mbox{GeV}^2\;,
\label{kg}
\end{equation}
so that
$$
\alpha_s = \alpha_s^{\overline{\rm MS}}(-e^{-5/3}k^2_g)\;.
$$
As one can see from Eq.(\ref{kg}), the virtuality of hard gluon is comparable
with the square of charm quark mass, and it indicates the applicability of the
hard process factorization. Moreover, the scheme-independent value of the
$\alpha_s$ argument $e^{-C_{scheme}}k^2/\Lambda^2_{scheme}$ is quite large, and
it is close to $15$. Numerically, the BLM fixing of the QCD coupling constant
gives $\alpha_s^{BLM}\approx 0.57$. In the above procedure the gluon
virtualities are taken into account. However, the quark virtualities can be
essential, too. Indeed, the charmed quark virtuality is valued in the
intermediate region, where it is less than the $b$-quark one and greater than
the gluon virtuality. So, in the following estimates we will use the value
$$
\alpha_s = \alpha_s^{\overline{\rm MS}}(m_c^2-k_c^2)\approx
\alpha_s^{\overline{\rm MS}}(-2 k_g^2) = 0.33\pm 0.06\;.
$$
The uncertainty in the $\alpha_s$ value appropriate for the given process 
indicates a possible large role of higher order corrections.

In numerical estimates we suppose \cite{neu}
$$
|V_{bc}| = 0.041\pm 0.003\;,
$$  
and we use the one-loop expression for the $\alpha_s$ evolution
$$
\alpha_s(m^2) = \frac{4\pi}{\beta_0(n_f)\ln(m^2/\Lambda^2_{(n_f)})}\;,
$$
where $\beta_0(n_f)=11-2n_f/3$, $n_f$ is the number of quark flavors with
$m_{n_f}<m$,
$$
\Lambda_{(n_f)} = \Lambda_{(n_f+1)} \biggl(\frac{m_{n_f+1}}{\Lambda_{(n_f+1)}}
\biggr)^{\frac{2}{3\beta_0(n_f)}}\;.
$$
Using $\alpha_s^{\overline{\rm MS}}(m_Z^2) = 0.117\pm 0.005$ \cite{pdg},
one finds  that $\Lambda_{(5)} = 85\pm 25$ MeV and $\Lambda_{(3)} =
140\pm 40$ MeV. 
One estimates $\alpha_s^{\overline{\rm MS}}(m_b^2) = 0.20\pm 0.02$, that is
quite reasonable.

The $f_{B_c}$ constant was estimated in the framework of the QCD sum rules
\cite{7,8,13}
$$
f_{B_c} = 385\pm 25\; \mbox{MeV,}
$$
and it is in a good agreement with the scaling relation for the leptonic
constants of $1S$ heavy quarkonia \cite{7}
$$
\frac{f^2}{M}\; \biggl(\frac{M}{\mu_{12}}\biggr)^2 = {\rm const.}\;,\;\;\;\;\;
\mu_{12} = \frac{m_1 m_2}{m_1+m_2}\;.
$$
Then, the account for the hard gluon corrections gives
\begin{eqnarray}
\tilde f_\psi = \tilde f_{\eta_c} & = & 542\pm 50\; \mbox{MeV,} \\
\tilde f_{B_c} & = & 440\pm 40\; \mbox{MeV.} 
\end{eqnarray}

To calculate the branching ratios we
evaluate the total $B_c$ meson width according to the formula by 
\cite{3,10}
$$
\Gamma(B_c) \approx \Gamma(B) + (0.6\pm 0.1) \Gamma(D^+) + \Gamma(ann.)\;,
$$
where $\Gamma(B)$ is the contribution of $\bar b$-quark decays with
the spectator $c$-quark, $\Gamma(D^+)$ determines the contribution 
of $c$-quark decays with the spectator $\bar b$-quark and with 
the account for the phase space reduction, because of the $c$-quark binding
inside $B_c$ (i.e. one takes into account the deviation from the exact
spectator consideration), and $\Gamma(ann.)$ is the contribution of 
annihilation channels depending on $|V_{bc}|$ and $f_{B_c}$. Then
$$
\Gamma(B_c) = (1.2\pm 0.2)\cdot 10^{-3}\; {\rm eV} = \frac{1}{0.55\pm 0.15\;
{\rm ps}}\;.
$$ 
The recent estimates of the $\Gamma(B_c)$ total width calculated within the 
operator product expansion approach and including also the bound quark
effects, annihilation channels as well as the Pauli interference in the final 
state of $B_c$ decay, is in a good agreement with the given value \cite{11}.

Supposing $a_1= 1.22\pm 0.04$ \cite{3}, $f_\rho = 220$ MeV, one finally finds
\begin{eqnarray}
\Gamma(B_c^+\to \psi \pi^+) &=& (9.2\pm 2.3)\cdot 10^{-6}\; {\rm eV} = 
\frac{1}{69\pm 17\; {\rm ps}}\;, \label{gam1-}\\
\Gamma(B_c^+\to \psi \rho^+) &=& (24\pm 6)\cdot 10^{-6}\; {\rm eV} = 
\frac{1}{22\pm 5\; {\rm ps}}\;, \label{gam1}
\end{eqnarray}
and
\begin{eqnarray}
{\rm BR}^{\rm HS}(B_c^+\to \psi \pi^+) &=& 0.77\pm 0.19 \%\;, 
~~~~~~~
{\rm BR}^{\rm HS}(B_c^+\to \eta_c \pi^+) = 1.00\pm 0.25 \%\;,
\label{br-}\\
{\rm BR}^{\rm HS}(B_c^+\to \psi \rho^+) &=& 2.25\pm 0.56 \%\;, 
~~~~~~~
{\rm BR}^{\rm HS}(B_c^+\to \eta_c \rho^+) = 2.78\pm 0.70 \%\;.
\label{br}
\end{eqnarray}

Further, the purely spectator decays of $\bar b \to \bar c \pi^+(\rho^+)$ 
have the following branching fractions with respect to the total $B_c^+$ width
\begin{eqnarray}
{\rm BR}^{\rm spec}(\bar b \to \bar c \pi^+) &\approx & 0.64\%\;, 
\label{spec1-}\\
{\rm BR}^{\rm spec}(\bar b \to \bar c \rho^+) &\approx & 1.8\%\;, 
\label{spec1}
\end{eqnarray}

The matrix element, corresponding to the first diagram in Fig.~1, is 
approximately equal to the matrix element, following from the second 
diagram and, hence, estimates (\ref{br-},\ref{br}) are enhanced by a 
factor of four  due to the $t$-exchange nonspectator contribution. 

As for the $nS$-excitation yields of the $(\bar c c)$ quarkonium in the $B_c^+$
decays, we note that the corresponding branching fractions are determined by 
the rescaling of the leptonic constants and phase spaces. So,
\begin{equation}
{\rm BR}^{\rm HS}(B_c^+\to \psi(nS) \pi^+(\rho^+)) =
{\rm BR}^{\rm HS}(B_c^+\to \psi \pi^+(\rho^+))\; \frac{f^2_{nS}}{f^2_\psi}\;
\frac{M^2-m^2_{nS}}{M^2-m^2_\psi}\;.
\end{equation}
The experimental values of leptonic constants are in a good agreement
with the scaling expression \cite{7}
\begin{equation}
\frac{f^2_{nS}}{f^2_\psi}= \frac{1}{n}\; \frac{m_\psi}{m_{nS}}\;.
\end{equation}
So, neglecting the differences in the masses of $\psi(nS)$ and $\eta_c(nS)$
states, one gets
\begin{equation}
\frac{{\rm BR}^{\rm HS}(B_c^+\to \psi(2S) \pi^+)}
{{\rm BR}^{\rm HS}(B_c^+\to \psi \pi^+)} =
\frac{{\rm BR}^{\rm HS}(B_c^+\to \eta_c(2S) \pi^+)}
{{\rm BR}^{\rm HS}(B_c^+\to \eta_c \pi^+) } \approx 0.36\;,
\end{equation}
for instance. The same values for the $\psi(nS)\rho^+$ state 
yields can be rewritten down.

\section{Conclusion}

In this paper we have shown that in the $B_c^+\to \psi \pi^+(\rho^+)$
and $B_c^+\to \eta_c \pi^+(\rho^+)$ decays
the large momentum of the recoil $\psi$ or $\eta_c$ particle leads to the fact 
that the formalism
of the weak transition form-factor calculation, based on the overlapping
of the nonrelativistic wave functions for the heavy quarkonia, is not valid.
The hard gluon exchange with the
spectator quark results in the large virtuality of heavy quark in the 
weak transition current. The amplitude of the weak 
decay with the hard exchange by gluon can be calculated in the framework of
QCD perturbation theory and this amplitude can be factorized from 
the amplitude of soft 
binding of heavy quarks in the quarkonium. The calculations with the account 
for this hard-soft factorization result in
\begin{eqnarray}
{\rm BR}(B_c^+\to \psi \pi^+) &=& 0.77\pm 0.19 \%\;, \nonumber\\
{\rm BR}(B_c^+\to \psi \rho^+) &\approx & 2.78\cdot 
{\rm BR}(B_c^+\to \psi \pi^+)\;, \nonumber\\
{\rm BR}(B_c^+\to \eta_c \pi^+(\rho^+)) &\approx & 1.28\cdot
{\rm BR}(B_c^+\to \psi \pi^+(\rho^+))\;,\nonumber
\end{eqnarray}
where the accuracy is basically restricted by uncertainties in the evolution
scale of the "running" QCD coupling constant, in the $c$-quark mass, and
the total $B_c$ width. The given estimate for the branching ratio of the
$B_c^+\to \psi \pi^+$ decay mode is significantly larger than the
extrapolation results by the potential models. This value strongly 
enhance the probability of $B_c$ observation in the current Tevatron and LEP
experiments with vertex detectors.

The author thanks A.Razumov and S.Slabospitsky for the software help in 
the figure creation and A.A.Likhoded for the valuable remarks.

This work is in part supported by the Russian Foundation of Fundamental 
Researches, grant 96-02-18216, and by the program "Russian State Stipends 
for young scientists".

\end{document}